# Towards Collective Superintelligence, a pilot study

Louis Rosenberg
Unanimous AI
Pismo Beach, California
Louis@Unanimous.ai

Gregg Willcox
Unanimous AI
Seattle, Washington
Gregg@Unanimous.ai

Hans Schumann
Unanimous AI
San Francisco, California
Hans@Unanimous.ai

*Abstract—* Conversational Swarm Intelligence (CSI) is a new technology that enables human groups of potentially any size to hold real-time deliberative conversations online. Modeled on the dynamics of biological swarms, CSI aims to optimize group insights and amplify group intelligence. It uses Large Language Models (LLMs) in a novel framework to structure large-scale conversations, combining the benefits of small-group deliberative reasoning and large-group collective intelligence. In this study, a group of 241 real-time participants were asked to estimate the number of gumballs in a jar by looking at a photo. In one test case, individual participants entered their estimation in a standard survey. In another test case, participants converged on groupwise estimates collaboratively using a prototype CSI text-chat platform called Thinkscape. The results show that when using CSI, the group of 241 participants estimated within 12% of the correct answer, which was significantly more accurate ($p<0.001$) than the average individual (mean error of 55%) and the survey-based Wisdom of Crowd (error of 25%). The group using CSI was also more accurate than an estimate generated by GPT 4 (error of 42%). This suggests that CSI is a viable method for enabling large, networked groups to hold coherent real-time deliberative conversations that amplify collective intelligence. Because this technology is scalable, it could provide a possible pathway towards building a general-purpose Collective Superintelligence (CSi).

## I. Introduction.

Collective Intelligence (CI) refers to the field of study that investigates how human groups can solve problems and make decisions that exceed the capabilities of the individual members. Often referred to as "wisdom of crowds," the most common CI methods involve capturing asynchronous input from large groups through polls, surveys, or prediction markets, and then aggregating the data statistically. This often produces more accurate results than the median individual produces on their own, thereby exhibiting measurable intelligence amplification. Such statistical techniques are commonly used in groupwise estimation as well as probabilistic forecasting and other simple decision-making tasks. Applicability is generally limited to tasks that can be abstracted down to discrete questions in which participants are asked to provide numerical input or make multiple-choice selections or forced-choice comparisons.

While CI technologies have migrated online and employ increasingly sophisticated statistics, often supplemented by simple machine learning, the basic process of asynchronous data collection has remained largely unchanged for over a century. It all dates back to 1907 and Sir Francis Galton's publication of "Vox Populi" in the journal Nature [1]. The short paper details his famous experiment at a livestock fair in which he collected estimations from 787 people for the weight of an ox and demonstrated that the median estimation was significantly more accurate than the vast majority of individual members. This process is often replicated for illustrative purposes using items that are more accessible than a live ox – for example, the number of jellybeans in a jar. In fact, the pilot study below uses gumballs in a jar and compares a traditional asynchronous survey method for groupwise estimation with a new synchronous CI technology known as Conversational Swarm Intelligence (CSI). It's based on the biological principle of Swarm Intelligence but is applied to real-time networked human groups using a unique application of Large Language Model (LLM) technologies.

## II. Swarm Intelligence (SI)

Over hundreds of millions of years, many biological species have evolved the ability to amplify their collective intelligence, making groupwise decisions that are significantly more accurate than individuals could make on their own. Biologists refer to the emergent decision-making abilities of natural organisms as *Swarm Intelligence* and it operates very differently from the statistical method pioneered by Galton. Instead of aggregating asynchronous estimates like humans do, biological populations form real-time synchronous systems that enable participants to engage in a multi-directional tug-of-war, pushing and pulling on the system until a solution emerges that best represents the combined sentiments of the population. This is why fish school, birds flock, and bees swarm – they can make critical decisions without any individual being in charge and those decisions can show intelligence levels that greatly exceed the capacity of their individual brains. From this perspective, nature points us not just to collective intelligence but Collective Superintelligence (CSi) and suggests the process requires the creation of synchronous systems, not asynchronous data collection and aggregation.

The most famous species for Swarm Intelligence in nature is likely the honeybee which has been studied for decades and arguably demonstrates superintelligent abilities [2,3]. For example, honeybees face a life-or-death decision when selecting a new location for their colony. They solve this by sending hundreds of scout bees into the world to search a 20 to 30 square mile area. The scouts investigate dozens of potential sites and bring their assessments back to the swarm for consideration. Picking the best site is a complex multivariable problem that involves many competing constraints. For example, the bees need to select a site that is large enough to store the honey they need for winter, close to good sources of pollen, safe from predators, insulated enough for winter, ventilated enough for summer, and has a nearby source of water.

A human (with approximately 86 billion neurons) would have difficulty selecting the very best option across the many completing constraints without computational support. And yet honeybees (each with less than 1 million neurons) have been shown to collectively solve this multi-variable problem. They do it by forming a real-time system in which scout bees express their preference through body vibrations known as a "waggle dance." Through a dynamic negotiation among the competing signals, a single decision emerges upon that maximizes support from scouts and it's usually the best possible decision. In this way, Swarm Intelligence is not asynchronous aggregation but a real-time interactive system that deliberates and converges.

In 2015 a new CI technology was proposed called Artificial Swarm Intelligence (or Swarm AI) to enable networked human groups to form real-time systems with features similar to bee swarms [4]. Because humans did not evolve the ability to waggle dance like bees, artificial swarms were created by enabling users to modulate real-time signals using their mice, touchscreens or trackpads [5]. This was done by giving each user the ability to control a graphical magnet such that hundreds of simultaneous users could engage in a tug-of-war in that collectively guides a graphical puck to solutions the group can best agree upon. In addition, real-time intelligence algorithms were developed to process the dynamic behaviors of individuals, converting their time-varying actions into inferred sentiment strengths. In this way, Artificial Swarm Intelligence is a combination of unique human interface and artificial intelligence technologies [6-8].

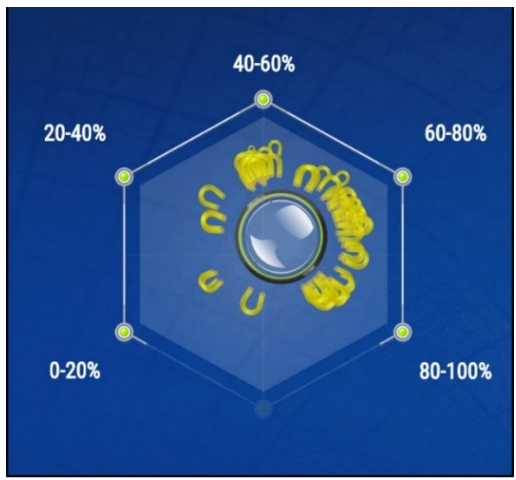

Fig. 1. Snapshot of Artificial Swarm Intelligence system of 100 participants

In recent years, many studies have been conducted that validate the ability of Artificial Swarm Intelligence to significantly outperform asynchronous collective intelligence methods. For example, Researchers at MIT and Unanimous AI conducted a study showing that groups of financial traders, when deliberating as real-time swarms, amplified forecasting accuracy by 36% ($p<0.001$) [9]. Similarly, researchers at Stanford University School of Medicine and Unanimous AI conducted an NSF-funded study showing that groups of doctors working in swarms reduced their diagnostic errors by 33% as compared to traditional methods [7, 10]. In addition, researchers at California Polytechnic State University (Cal Poly) demonstrated that distributed human teams could significantly amplify their "social sensitivity" and "social perceptiveness" when deliberating in real-time swarms [11, 12].

As shown in Figure 1, current Artificial Swarm Intelligence techniques allow human groups deliberate among predefined sets of options and collectively rate, rank, or select among them [6, 8]. While these capabilities are valuable in a wide range of commercial applications from forecasting and diagnosis to prioritizing objectives, current methods are too narrow to lead to a general-purpose Collective Superintelligence. That's because human participants cannot adequately express their individual knowledge, wisdom, insights, or intuition when responding to questions through numerical forecasts, subjective comparisons, or multiple-choice selections. To enable a pathway to a true CSi, the biologically inspired concept of dynamic swarms needs to be reconciled with the core human ability to deliberate through natural language. To address this, a *conversational collective intelligence* technology was introduced in 2023 known as Conversational Swarm Intelligence and it aims to amplify the combined intellect of large networked human groups through real-time conversational deliberation [13].

### III. CONVERSATIONAL SWARM INTELLIGENCE (CSI)

The motivation for developing the unique CSI architecture was to enable very large human groups to engage in real-time conversational deliberations and converge on optimal solutions that maximize group satisfaction, conviction, and accuracy [14]. This poses a unique challenge as there are no communication methods currently available to enable real-time conversational deliberation among large networked groups. For example, you could bring hundreds of people into a single chatroom, as is often done on Twitch and other large real-time platforms, but that does not yield meaningful dialog or coherent deliberation. That's because conversational quality degrades with group size [15]. Sometimes referred to as the "many minds problem," when groups grow beyond a handful of people, the conversational dynamics fall apart, providing less "airtime" per person, disrupting turn-taking dynamics, providing less feedback per comment, and reducing engagement as participants feel less social pressure to participate. In fact, putting dozens of users in a single chatroom or videoconference would not yield an authentic "conversation," but would instead devolve into a stream of singular comments with little interaction among them.

To address this impasse, researchers took inspiration from the swarming behaviors of large fish schools [16]. Even with thousands of members, large schools can hold coherent real-time "conversations" among the full population, enabling rapid groupwise decisions without any central authority mediating the process. Each fish communicates with others using a unique organ called a "lateral line" that senses pressure changes caused by neighboring fish as they adjust speed and direction with varying levels of conviction. The number of neighbors that a given fish pays attention to varies from species to species, but it's always a small subset of the group. And because each fish reacts to an overlapping subset of other fish, information quickly propagates across the full population, enabling a single Swarm Intelligence to emerge that rapidly converges on unified decisions [17, 18].

Conversational Swarm Intelligence follows a similar model to fish schools, enabling large, networked groups to hold real-time conversations such that each participant can hold a coherent and thoughtful conversation with a small subset of their neighbors while also allowing conversational content to freely propagate across the full population. Using this technology, a networked group of 400 users could be divided into a large number of smaller subgroups, for example 80 groups of 5 people, with the members of each subgroup routed into their own chat room or video conference and tasked with discussing an issue in parallel with the other 79 subgroups. Of course, simply subdividing the population into 80 subgroups will not yield a conversational collective intelligence unless information unless information can propagate across the population. Researchers solved this issue through the novel use of AI agents powered by Large Language Models (LLMs) to emulate the functionality of the lateral line organ in fish [13, 14, 16].

To achieve this, an AI-powered "Observer Agent" is inserted into each of the 80 rooms and tasked with observing the deliberative dialog in that room, distilling salient content and expressing it in a neighboring room as first-person dialog. In this way, each of the 80 groups is given an additional member that is an AI observer tasked with expressing the insights that emerge in one group into neighboring groups. This enables information to propagate across the full population of 400 people in a natural and seamless way that feels like a focused deliberation. This creates a single system in which 250 or 2,500 or even 25,000 people could hold a real-time deliberation, sharing views and ideas, debating options, and converging on unified solutions that optimize overall support. An example CSI structure for 100 members is shown below in Figure 2.

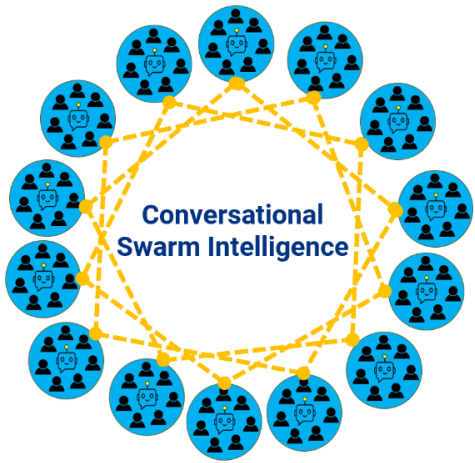

Fig. 2. Architecture for a Conversational Swarm with 100 real-time members

Conversational Swarm Intelligence has two major benefits over prior collective intelligence systems. First, it enables the use of open-ended questions in which answer options are not known in advance, empowering participants to suggest and debate an unlimited set of alternatives that are not pre-defined. And second, it allows users to not only express which options they personally prefer but also argue for or against options with specific reasons, rationales, or justifications. In this way, the CSI technology can not only facilitate convergence of large groups on unified solutions, but can also capture the reasoning for why groups supports the solutions they do. In addition, CSI is designed to mitigate social influence bias because each member is only influenced by a small number of others in real-time, reducing the impact of early views and/or strong personalities on the full population [17]. In this way, CSI was designed to combine the intelligence amplification benefits of large-scale polling with the qualitative benefits of small-scale deliberations. In addition, because CSI is a scalable technology that could be used to connect thousands or even millions of real-time users, it has the potential to enable super-intelligent capabilities.

In a prior CSI study conducted at Carnegie Mellon, real-time groups of 25 participants using Thinkscape were compared to standard chat rooms. The tests showed that participants in the CSI structure produced 30% more contributions ($p<0.05$) than those using a standard chat room and 7.2% less contribution variance, indicating that users contributed more content and participated more evenly when using CSI [13].

In another prior CSI study, groups of 48 users were tasked with debating the risk of AI on society. The participants using CSI contributed 51% more content ($p<0.001$) compared to those using standard chat. In addition, CSI deliberations showed 37% less difference in contribution quantity between the most vocal and least vocal contributors, thereby indicating that CSI fosters more balanced deliberations. And finally, a large majority of the participants preferred CSI system over standard chat ($p<0.05$) and reported feeling more impactful ($p<0.01$) [14].

In a third prior study, a group of 80 participants was tested in Thinkscape to assess the ability of CSI systems to generate qualitative insights on controversial issues. In this case, the topic for discussion for the 80-person population was the Republican Primary for President of the United States in 2023. After only six minutes of chat-based deliberation, the group converged on a preferred candidate and surfaced over 200 reasons for supporting that candidate. In addition, statistically significant results emerged in only six minutes, where the selected candidate generated significantly more support than the other options among the 80 participants ($p<0.001$) [16,17].

While these prior studies have been insightful, they did not assess the ability of CSI to amplify group intelligence through real-time conversations. To measure the potential of CSI in this regard, a new pilot study was conducted using a group of 241 networked users engaged in a single unified conversation via text. The task given to the group was to estimate the number of gumballs in a jar, thereby emulating the famous Galton experiment described above. The CSI method was compared against a standard survey, thereby testing traditional collective intelligence against conversational collective intelligence.

IV. COLLECTIVE INTELLIGENCE STUDY

An online chat-based CSI platform called Thinkscape was used to conduct a subjective estimation task in which 241 real-time participants were asked to estimate the number of gumballs in a jar and to support their answer with reasoning. The group was given 4 minutes to deliberate which answer was closest to the correct answer.

The CSI software automatically partitioned the 241 real-time participants into 47 subgroups of 5 or 6 members, each sized to enable coherent local deliberations using text-chat. The CSI software also deployed an AI Observer Agent into each of the 47 groups, where each AI agent was tasked with repeatedly passing conversational information that was observed in its group into a neighboring group. This information was expressed by each AI agent as first-person dialog and was configured to represent the top choices deliberated in its group along with the top rationales in support those choices. This unique use of AI agents enables conversational overlap between neighboring groups throughout the 47-group structure. In this way, real-time conversational information could propagate across the full population, helping the group have a single unified conversation. In this way, the CSI architecture enabled 47 thoughtful local deliberations while also enabling the benefits of large-scale global collective intelligence.

The CSI process was compared to a traditional survey-based estimation. In both cases, the 241 participants were given up to 4 minutes to formulate their estimations. In both cases the participants were shown a photograph of the jar filled with gumballs as shown in Figure 3. For both the survey and the CSI trials, the participants were given a set of ten possible answers to choose among and asked to pick the best option. The photo used was adapted from a prior CI research effort in which groups of human participants were individually polled for numerical estimations [28]. The answer choices used in the current study were generated by sampling the response distribution in the prior study. The same set was used for both trials.

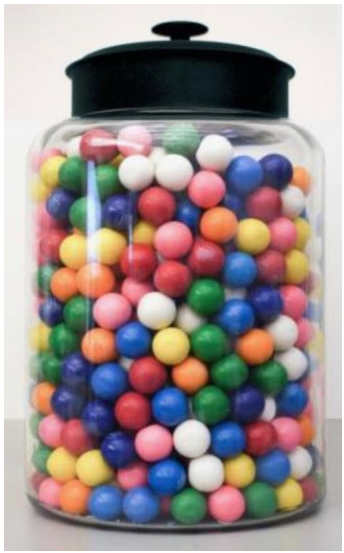

Fig 3. Image of gumballs in a jar.

The survey-based estimations were made asynchronously with each user working on their own and the data aggregated statistically at the end. The conversational method using CSI employed natural language processing to assess the strength of conversational support for each of the ten answer options at regular intervals during the 4-minute deliberation period. The system also computed a weighted average across the ten options based on the conversational sentiment levels. This weighted average was not shown to participants in real-time but was used to compute a final numerical estimation for conversational group. Figure 4 below shows the evolving sentiments in favor of each answer option over the 4-minute period and the final weighted average produced by the CSI system.

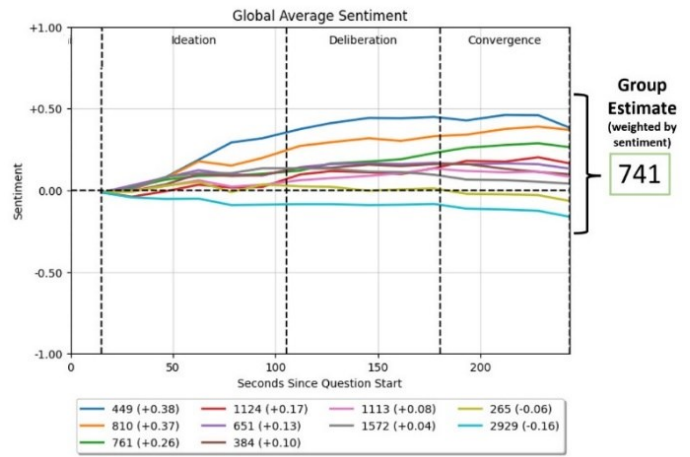

Fig 4. Conversational Sentiment for each answer option during 4-min deliberation in Thinkscape.

In addition, GPT-4.0 was given the same photograph of the jar and asked to estimate the number of gumballs inside. It provided a detailed answer as follows:

*Estimating the number of gumballs in a jar is a fun and often challenging exercise. The best way to approach this is by breaking it down step-by-step:*

*Volume of the Jar: We need to approximate the volume of the jar. If it looks like a cylinder, the formula is $\pi r^2 h$, where r is the radius and h is the height. Without dimensions, we have to make an assumption based on the image.*

*Volume of a Gumball: Gumballs are roughly spherical, so the volume of a single gumball can be estimated using the formula for the volume of a sphere: $4/3 \pi r^3$. Again, without exact dimensions, we have to make an assumption on the size of a typical gumball.*

*Calculate Number of Gumballs: By dividing the volume of the jar by the volume of a single gumball, we can get an approximate number of gumballs the jar can contain. This doesn't account for the space between gumballs, so the real number is likely a bit less than this estimate.*

*Using the image alone, without measurements, it's difficult to provide an exact number. However, as a rough estimate:*

*Assuming:*

*A typical gumball has a diameter of about 2 cm.*

*The jar has a diameter of 15 cm and a height of 25 cm.*

*Volume of gumball $\approx 4/3 \pi (1)^3 = 4.19$ cm^3 (rounded)*

*Volume of jar $\approx \pi (7.5^2)(25) = 1767.15$ cm^3 (rounded)*

*Number of gumballs = 1767.15/4.19 $\approx$ 421 gumballs.*

*Accounting for spaces between gumballs, the number might be closer to **370-390** gumballs.*

*Again, this is a very rough estimate based on assumptions and the image provided. The most accurate way would be to measure the jar and a gumball or to physically count them.*

As is often the case, ChatGPT expresses confidence in its methodology but also hedges its forecast, providing a range of 370 to 390 gumballs and indicating that it can only give a very rough estimate from a photo. Of course, the humans in this pilot test had the same constraint. Ultimately, the big question is, which method was more accurate: (a) individual humans, (b) the statistical mean across 241 humans, (c) a group of 241 humans using Conversational Swarm Intelligence, or (d) GPT 4.0 using the same photo and making its best estimate.

## V. Results

Within the allotted four minutes, the real-time deliberative conversation across 241 networked participants…

i. **Individual Humans** (survey): computing the mean absolute error across the population of 241 participants who provided answers on the survey, we find that the average human was off by 361 gumballs (55%).

ii. **AI Estimate** (ChatGPT 4.0): computing the mean absolute error based on the center of the range (370 to 390) that was given by ChatGPT in its answer, we find that the AI system was off by 279 gumballs (42%)

iii. **Standard Collective Intelligence** (survey): computing the statistical mean across the 241 estimates made by survey respondents, we find that the Wisdom of the Crowd (WoC) was off by 163 gumballs (25%)

iv. **Conversational Swarm Intelligence** (CSI): when the 241 participants used Thinkscape and deliberated the number of gumballs conversationally, the weighted mean of human input (scaled by conversational sentiment) was off by 82 gumballs (12%).

The results summarized above and shown in Figure 5 below reveal a variety of interesting observations. First, we see that a pure AI system (GPT 4.0) given the photograph was able to perform better than the average individual human on this task, recording a 23% smaller error. This is an impressive result and points towards the evolution of superintelligent AI systems.

Second, it showed that using a standard survey-based collective intelligence technique, the human group of 241 participants was able to outperform GPT 4.0. Using the statistical mean of 241 human estimates, the human group achieved 42% smaller error in estimation than GPT 4.0. This suggests that collective intelligence is also a viable path to amplified intelligence.

Third, the experimental CSI method in which the 241-member group was structured into 47 deliberative subgroups and used AI agents to pass conversational information between groups was significantly more accurate than all three of the other test cases. In particular, the CSI method outperformed the average individual, recording 77% smaller estimation error. Using a one-tailed z-test, CSI was found to significantly outperform the average individual ($p<0.001$). This is a strong indication that CSI is a viable method for amplifying the collective intelligence of networked human groups. In addition, CSI technology outperformed the GPT 4.0 estimation with 70% smaller error. This suggests that CSI may be a viable path for large human groups to conversationally form super-intelligent systems that can deliberate problems in real-time.

Finally, Conversational Swarm Intelligence produced a far more accurate results than the traditional survey-based CI technique, achieving a 50% lower estimation error. It is promising that CSI was more accurate than the survey mean. This suggests that CSI is not only a more flexible and insightful CI method than traditional surveys (by enabling free form conversations), it may also be a more accurate technique as well.

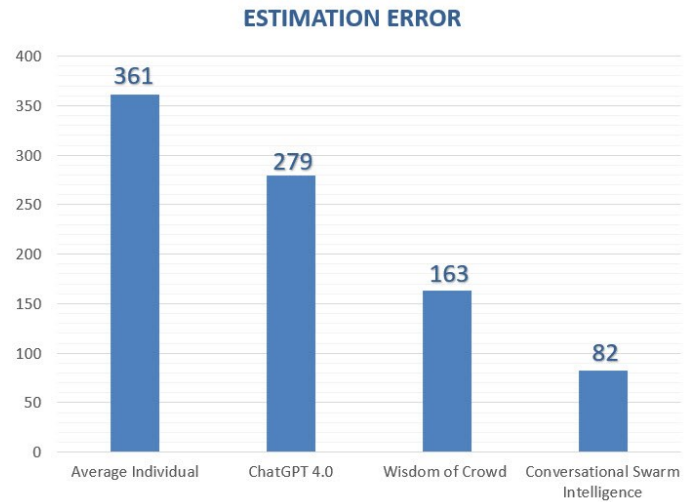

Fig 5. Comparison of estimation error across four test methods.

## VI. Conclusions

We conducted a collaborative estimation task using a new technology called Conversational Swarm Intelligence. It combines the methods inspired by the biological principle of Swarm Intelligence with a novel use of LLMs that enables large, distributed groups to hold coherent conversational deliberation and reach thoughtful and unified solutions. We performed a study with a large 241-member group of randomly selected users and tested their ability to estimate the number of gumballs in a jar. This task was chosen as it replicates a very common test used to illustrate the power of collective intelligence. The results showed that "conversational collective intelligence" is a viable method in which human groups deliberate through natural language and reach solutions of amplified accuracy. In this case, the use of CSI technology reduced the mean absolute error across individual participants by 77% ($p<0.001$). This was more effective than the traditional collective intelligence method of aggregating asynchronous survey results. In fact, the estimation error using CSI was 50% smaller than the error using the mean survey estimate. And finally, we found that although GPT 4 outperformed the average human in this estimation task, CSI enabled human group to form a conversational collective intelligence that outperformed GPT 4, reducing the estimation error by 70%. This suggests that CSI be a viable path towards building a real-time collective superintelligence that can harness real-time human knowledge, wisdom, insights, and intuition through natural conversational deliberations.

Future work will test larger groups, with milestones being real-time conversations among 1000 individuals or more. In addition, applications for Enterprise Collaboration, Consumer Intelligence, Political Intelligence, and Deliberative Civic Engagement are currently being tested among authentic groups.

In addition, applications using voice-chat, video conferencing, and VR chat are all anticipated within the CSI architecture and are being investigated in parallel to text chat. The goal of developing largescale CSI systems is to explore near-term value for amplifying the intelligence of organizational teams and long-term value for achieving Collective Superintelligence.

ACKNOWLEDGMENT

The authors thank Chris Hornbostel and Patty Sullivan for their efforts recruiting participants and moderating sessions.